\documentstyle[prb,twocolumn,aps,epsf]{revtex}

\begin{document}

\title{Molecular-field approach to the spin-Peierls transition
       in CuGeO$_3$}
\author{Ralph Werner and Claudius Gros}
\address{Universit\"at Dortmund, Institut f\"ur Physik, 
         D-44221 Dortmund, Germany}

\date{\today}
\maketitle
                    
\begin{abstract}
We present a theory for the spin-Peierls transition 
in CuGeO$_3$. We map the elementary excitations of the
dimerized chain (solitons) on an effective Ising model. Inter-chain
coupling (or phonons) then introduces a linear binding potential
between a pair of soliton and anti-soliton, leading to a finite
transition temperature. We evaluate, as a function of temperature,
the order parameter, the singlet-triplet gap, the specific heat,
and the susceptibility and compare with experimental data
on CuGeO$_3$. We find that CuGeO$_3$ is close to a first-order
phase transition. We point out, that
the famous scaling law $\sim\delta^{2/3}$ of the triplet gap
is a simple consequence of the linear binding potential between
pairs of solitons and anti-solitons in dimerized spin chains.
\end{abstract}

\section{Introduction}

With the discovery of the first inorganic spin-Peierls
compound CuGeO$_3$ \cite{Hase} it has become
possible to investigate the
physics of the spin-Peierls transition in quasi  
one-dimensional spin chains with high precision.
Of particular interest has been the study of
the magnetic excitation spectrum by neutron
scattering \cite{Nishi,Regnault,Martin} and
Raman scattering experiments
\cite{Loosdrecht,Muthu}. 
It has been observed \cite{Martin,Lussier},
that the spin-Peierls gap, i.e.~the gap to
triplet excitations out of the singlet ground-state, 
has a temperature dependence which
is difficult to explain with existing theories 
of the spin-Peierls transition \cite{Pytte,Cross}.

Existing theories of the spin-Peierls transition
of spin-1/2 Heisenberg chains are based on the mapping 
of the spin chain to an interacting gas of spin-less
fermions via the Jordan-Wigner transformation \cite{Pytte,Cross}.
The chains are coupled to a three-dimensional
phonon mode and the transition occurs when the respective
phonon frequency becomes soft. In the spin-Peierls state
the unit-cell doubles due to the alternating displacements
of the magnetic ions. The effective Hamiltonian for the
magnetic excitations along the chains
in the dimerized state is then \cite{Castilla}

\begin{equation}
H = J\sum_i\  \left[ (1+\delta(-1)^i)\,{\bf S}_i\cdot{\bf S}_{i+1}
            +    \alpha\,       {\bf S}_i\cdot{\bf S}_{i+2}
              \right].
\label{H}
\end{equation}
The nearest neighbor (NN) exchange has alternating 
strength $\ J(1\pm\delta)\ $ due to the doubling of the
unit cell below the spin-Peierls transition temperature $\ T_{SP}$.
Also included in (\ref{H}) is a frustrating next-nearest-neighbor
(NNN) term $\ \sim J\alpha$. The ground-state phase diagram of (\ref{H}) has
been mapped out by a density matrix renormalization group study \cite{Chitra}.
There is a gap everywhere in the phase-diagram except
on the line $\ \delta=0\ $ and $\ \alpha<\alpha_c$,
with $\ \alpha_c\approx 0.2411$. A vividly debated question
of key interest is therefore the actual magnitude of the
strength of the frustration $\ \alpha$. 
For CuGeO$_3$ the parameters have been estimated as
$\ J\approx150-160\mbox{K}\ $ and $\ \alpha\approx0.24-0.35$
\cite{Castilla,Fabricius,Riera}. 

A consequence of existing theories of the spin-Peierls transition
is the occurrence of a soft phonon mode above $\ T_{SP}$, which
has not yet been observed for CuGeO$_3$ \cite{Hirota}.  
It is therefore of interest to pursue
the possibility of a purely electronically driven
spin-Peierls transition, as it could occur for
$\ \alpha>\alpha_c$. In such a scenario the phonons would
just follow the pre-formed electronic dimerization, i.e.\ the 
formation of NN singlet pairs. Here we propose a simple
molecular-field type theory for the spin-Peierls transition
by mapping the low-energy solitonic excitations of
dimerized spin chains on an effective Ising model.
We emphasize that this procedure is valid both for the
phonon-driven and for the electronically-driven 
spin-Peierls transition. We present results of the
mean-field theory in comparison with experiments on CuGeO$_3$.
We find that a straightforward determination of the frustration
parameter $\ \alpha\ $ is not conclusive and that CuGeO$_3$ is
close to a first-order phase transition. We also note
that the famous $\ \sim\delta^{2/3}\ $ scaling of
the triplet gap \cite{Cross,Chitra} is a simple consequence
of the linear binding potential between solitons and
anti-solitons in dimerized spin chains.

\section{Mapping on an effective Ising model}

\begin{figure}[htb]
\epsfxsize=0.48\textwidth
\centerline{\epsffile{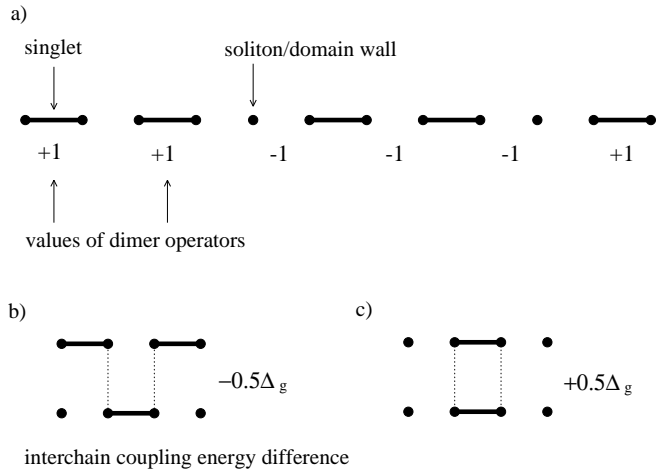}}
\medskip\medskip\medskip
\centerline{\parbox{0.48\textwidth}{\caption{\label{dimerchain}\sl
Phenomenological picture of copper chains in CuGeO$_3$.
a): A soliton and an anti-soliton in a dimerized
chain with the corresponding values of the dimer operators
$\sigma_{i,l}$. There is one dimer operator for every pair of
sites. b) and c): Illustration of two different dimer configurations
which lead to an inter-chain coupling contribution to the energy of 
$\,\pm0.5\Delta_g$. The origin of the coupling may be attributed to the
inter-chain exchange as well as to phononic effects.}}}
\end{figure}

A strictly one-dimensional system shows no long-range order at finite
temperatures due to the solitonic excitations. The appropriate
solitons in a spin-Peierls state are domain-walls in between
two different dimer coverings, see Fig.\ \ref{dimerchain}.
In this picture a dimer consists of a NN pair of spins in singlet state.
A single soliton in an otherwise (dimer-) ordered
chain is in reality a complicated object \cite{Shastry}.
It is spatially extended \cite{Keimer} and has a
spin degree of freedom together with a dispersion, which
in the Majumdar-Gosh model ($\ \alpha=1/2\ $ \cite{Gosh})
has the form $\ J\,(5/4-\cos 2 k)$.
It has been shown \cite{Shastry}, that the solitons yield an 
accurate description of the magnetic susceptibility 
and hence of the Hilbert space.

The dispersion of the solitons might be approximated, in general, by

\begin{equation}
E(k)\ =\ \sqrt{(E_s)^2+(c\,J\,\sin k)^2},
\label{dispersion}
\end{equation}
where $\ E_s\ $ is the gap to solitonic excitations, 
$\ J\ $ the exchange integral, $\ c\ $ a constant
of order of unity, and $\ k\ $ is the wavenumber in units of 
inverse lattice spacings. Only the low-energy solitons are
effective in destroying the long-range dimer order,
and we approximate here (\ref{dispersion}) by a 
constant $\ E(k)\equiv E_s$. We will furthermore not
discuss the consequence of the spatial extent of the solitons 
and we will include their spin-degree of freedom only when determining
the susceptibility.
We take one of the two possible dimer configurations of the
linear chain as the reference state and label every
``good'' dimer by $\ +1\ $ and every bond between two
``wrong'' dimers by $\ -1$; A bond between a soliton
and a wrong dimer is also labeled $\ -1$ as shown in 
Fig.\ \ref{dimerchain}.
Every possible soliton configuration is therefore mapped
to a Ising-spin configuration living on every second bond.
A similar mapping has been used recently by Mostovoy and Khomskii
\cite{Khomskii}.

We now consider a two-dimensional array of chains and define by
$\ \sigma_{i,l}=\pm1\ $ the Ising variable on the 
$l^{\mbox{\small th}}$ chain,
where the site index $\ i\ $ runs over every second bond of a chain
of length $\ L$. Including a coupling $\ \Delta_g\ $ 
between dimers on NN chains we have

\begin{equation}
H_{2D}\ =\ E_s\sum_{i,l}\frac{1}{2}(1-\sigma_{i,l}\sigma_{i+1,l})-
     \frac{\Delta_g}{2}\sum_{i,l}\sigma_{i,l}\sigma_{i,l+1}
\end{equation}
as the effective Ising Hamiltonian. The inter-chain coupling
$\ \Delta_g\ $ might be induced by the inter-chain exchange
coupling, $\ J_{\perp}\approx0.1 J$. In this case
$\ \Delta_g\sim (J_\perp)^2/J$. An indirect coupling of the chains via
phonons is also conceivable \cite{Pytte}. It has
been shown recently, that a molecular-field 
decoupling of inter-chain interactions in
quasi one-dimensional systems is a good approximation
in the strongly anisotropic limit \cite{quasi-1D}.
We may therefore decouple the chains via

\begin{equation}
\sigma_{i,l}\sigma_{i,l+1}\ \rightarrow\ 
\sigma_{i,l}\left\langle \sigma_{i,l+1} \right\rangle 
     + \sigma_{i,l+1}\left\langle \sigma_{i,l} \right\rangle 
     - \left\langle \sigma_{i,l} \right\rangle 
          \left\langle \sigma_{i,l+1} \right\rangle,
\end{equation}
where the $\ \langle\sigma_{i,l}\rangle\ $ denotes the
thermodynamic expectation value.
Translational invariance perpendicular to the chains
yields $\ \langle\sigma_{i,l+1}\rangle =
\langle\sigma_{i,l}\rangle =\langle\sigma\rangle\ $ and the
Hamiltonian for a single chain (the number of Ising variables is
$L/2$) becomes in this mean-field approximation 

\begin{equation}
H\ =\ -\frac{1}{2}E_s\sum_{i=1}^{L/2}\sigma_{i}\sigma_{i+1}-
     B\sum_{i=1}^{L/2}\sigma_{i}
     +\frac{L}{4}E_s+\frac{L}{4}\Delta_g\langle\sigma\rangle^2,
\label{Ising}
\end{equation}
where we have set $\ B=\Delta_g\langle\sigma\rangle$. 
This is just the Ising Hamiltonian for a ferromagnetic spin chain
in an external magnetic field $\ B$.
Alternatively we can interpret (\ref{Ising}) as an effective model 
for domain walls in a spin-Peierls
state with a linear binding potential 
$\ V(x)=B\cdot(x+1)$ (with $\ x\ \ge\ 1\ $ being the distance in sites,
not bonds) between a soliton and an anti-soliton.

So far we have neglected the spin-phonon coupling $\ \lambda_{sp}$.
A finite value for the spin-singlet order parameter
$\ \langle\sigma\rangle\ $
leads through $\ \lambda_{sp}\ $ to a finite lattice
dimerization and consequently to a modulation of the exchange 
constant, $J(1\pm\delta)$, in (\ref{H}), with
$\ \delta\sim\lambda_{sp}\langle\sigma\rangle$. The
spin-phonon coupling therefore adds
a term $\ \sim \lambda_{sp}\langle\sigma\rangle\ $ to the
confining potential $\ B$. On a mean-field level this
corresponds to a rescaling of the coupling constant

\begin{equation}
\Delta_g\ \rightarrow\ \Delta_g + c_1\lambda_{sp},\qquad\quad
B\ \rightarrow\ B + c_2\delta,
\label{Delta_g}
\end{equation}
with appropriate constants $\ c_1\ $ and $\ c_2$.

The partition function $Z$ can be obtained from (\ref{Ising}) by the
transfer matrix method. The free energy $\ F=-k_BT\ln Z\ $ is given in
the thermodynamic limit ($L\rightarrow \infty$) by 

\begin{equation}
{F\over L/2}\ =\ -k_BT\ln\lambda_0
   +\frac{1}{2}\Delta_g\langle\sigma\rangle^2
\label{free}
\end{equation}
where 
\begin{equation}
\lambda_0\ =\ \cosh(\beta\Delta_g\langle\sigma\rangle)+
\sqrt{\sinh^2(\beta\Delta_g\langle\sigma\rangle)
     +{\rm e}^{-2\beta E_s}}
\label{lambda_0}
\end{equation}
with $\ \beta=1/(k_B T)\ $ being the inverse temperature.
From the free energy all physical quantities can be derived.

\section{Excitation energies\label{sec_ex}}

We have two relevant excitation energies, the gap to
single soliton excitations, given by $\ E_s\ $ in
(\ref{Ising}), and the gap to triplet excitations, $ \Delta_N $,
as measured in a neutron scattering experiment \cite{Martin}.
Let us now discuss the relation of $\ \Delta_N$ to $\ E_s$.

A triplet excitation can dissolve into a soliton/anti-soliton pair. The
linear binding energy $\ V(x)=B\cdot(x+1)$ in between a soliton and 
an anti-soliton in (\ref{Ising}) leads to a confinement of
soliton/anti-soliton pairs \cite{Khomskii2}. In (\ref{Ising}) we have
neglected the kinetic energy of the single solitons
(\ref{dispersion}), which is of order $\ J$. The energy levels of a
particle in a linear confining potential are well known
\cite{Potential,Bulaevskii}. The lowest eigenstate has, in
the limit $\ B\ll J$, the energy 

\begin{equation}
\Delta_N\ =\ 2E_s + c^{\prime}J \left({B\over J}\right)^{2/3},
\label{Delta_N}
\end{equation}
with $\ c^{\prime}\approx2.33$. Eq.\ (\ref{Delta_N})
gives then the gap to triplet excitations as a function
of soliton energy $\ E_s\ $ and the strength of the
confining potential $\ B$.
The mean extension of a soliton/anti-soliton pair scales like
$\ (B/J)^{-1/3}$. 

For $\ \alpha<\alpha_c\ $ we have $\ B\rightarrow c_2\delta$ and
Eq.\ (\ref{Delta_N}) takes then the form

\begin{equation}
\Delta_N\Big|_{\alpha<\alpha_c} \
\approx\ 2E_s + 2.33\,v_s \left({c_2\delta\over v_s}\right)^{2/3},
\label{E(delta)}
\end{equation}
where $\ v_s\approx(\pi/2)(1-1.12\alpha)J\ $ is the 
$\alpha$-dependent spin-wave
velocity \cite{Fledder}. It is known \cite{Cross,Chitra}, 
that the gap scales for $\ \alpha<\alpha_c\ $ like
$\ \sim\delta^{2/3}\ $ implying that
$\ E_s\delta^{-2/3}\to0\ $ for $\ \delta\to0$.
Comparison with numerical results \cite{Chitra}
leads to $\ c_2\approx0.85$.
The functional form of the dependence of the
soliton excitation energy $\ E_s\ $ on $\ \delta$,
or alternatively on the
dimerization order parameter $\ \langle\sigma\rangle\ $
is not known at present.
It might be extracted for $\ \alpha<\alpha_c$,
in principle, from a sub-leading scaling
analysis of the excitation energy

\begin{equation}
\Delta_N\ =\ \tilde c\,J\,(\delta)^{2/3} 
            + 2 E_s(\delta),
\label{subleading}
\end{equation}
but this has not yet been done. We have therefore decided
to assume the functional form

\begin{equation}
E_s(\langle\sigma\rangle)\ =\
E_\infty+(E_0-E_\infty)\langle\sigma\rangle^2 ,
\label{E_s}
\end{equation}
where $\ E_0\ $ is the zero-temperature value of
$\ E_s\ $ and where $\ E_\infty\ $ is the soliton energy in 
the disordered phase, i.e.\ for $\ T>T_{SP}$.
Eq.\ (\ref{E_s}) is, in the spirit of a Landau
functional (\ref{Landau}), the simplest form
consistent with the symmetry of $\ \langle\sigma\rangle$.

\section{Landau expansion}

\begin{figure}[hbt]
\setlength{\unitlength}{0.96\textwidth}
\begin{picture}(0.49,0.80)
\epsfxsize=0.47\textwidth
\put(0,0.40){\epsffile{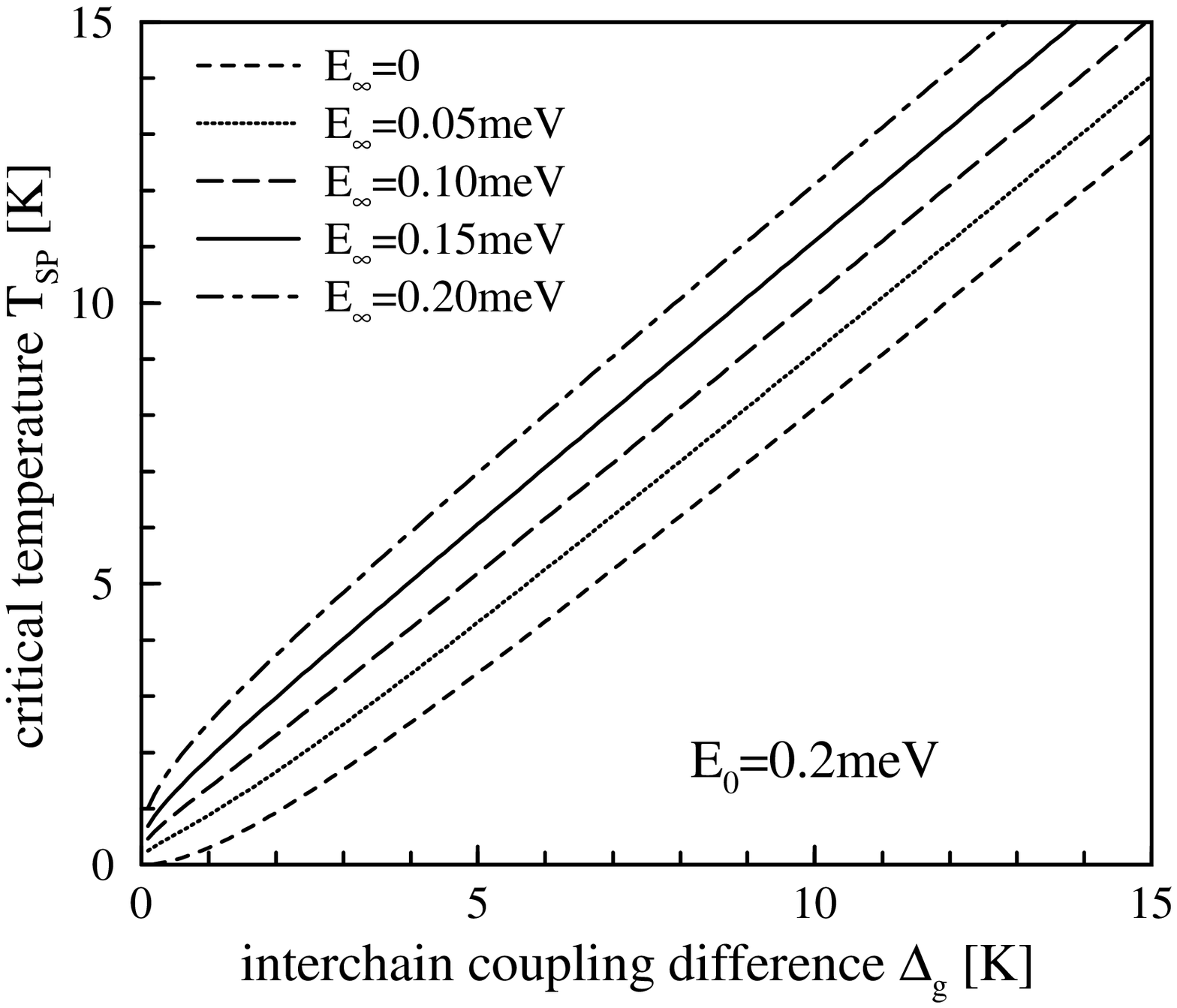}}
\epsfxsize=0.47\textwidth
\put(0,0){\epsffile{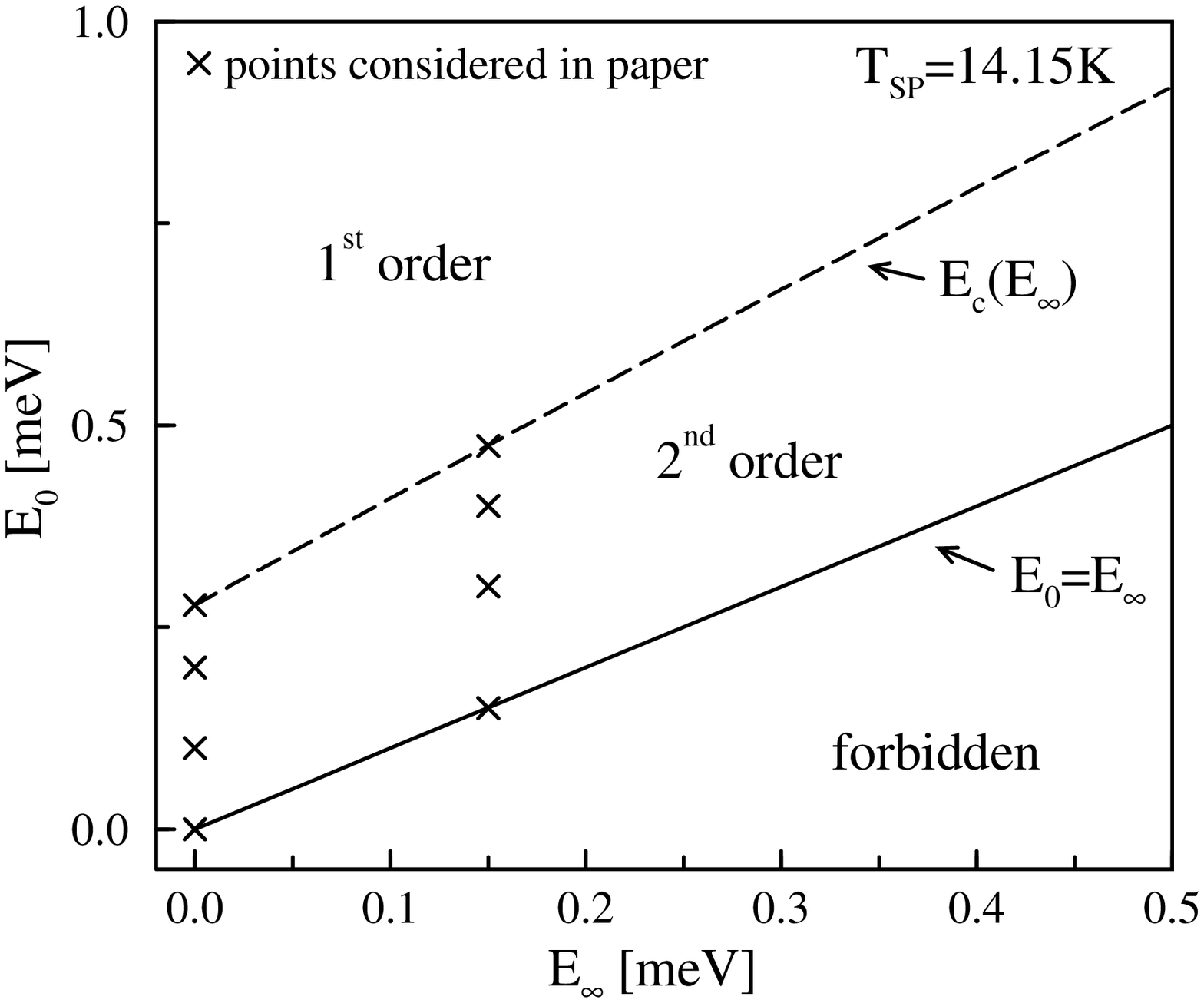}}
\put(0,0.42){(a)}
\put(0,0.02){(b)}
\end{picture}
\parbox{0.5\textwidth}{\caption{\label{phases} \sl (a):
Illustration of the spin-Peierls transition temperature 
$T_{SP}$ for $E_0=0.2\mbox{meV}$ as a function of $\Delta_g$.
(b):
Phase diagram as a function of $E_\infty$ and $E_0$ for fixed
$T_{SP}=14.15\mbox{K}$. Only the region $E_0>E_\infty$ is allowed
(above the solid line). Above the dashed line the phase transition
is of first-order, below it is of second-order. The crosses indicate
the parameter values considered for comparison with CuGeO$_3$.
}}
\end{figure}

We may expand the free energy (\ref{free}) in powers
of $\ \langle\sigma\rangle$,

\begin{equation}
F\ =\ F_0+a(T)\langle\sigma\rangle^2+b(T)\langle\sigma\rangle^4
+\mbox{O}(\langle\sigma\rangle^6).
\label{Landau}
\end{equation}
Depending on the parameters we may have either a 
second-order phase transition with $\ b(T) >0\ $ or a first-order
phase transition with $\ b(T)<0$. In the first case the
transition temperature $\ T_{SP}\ $ is given by
$\ a(T_{SP})=0\ $ as

\begin{eqnarray}
\lefteqn{
k_BT_{SP}\ =\ }\hspace{4ex}\nonumber\\&&
{\Delta_g^2(1+{\rm e}^{-\beta E_\infty}) \over
     2(E_0-E_\infty){\rm e}^{-2\beta E_\infty}
 +\Delta_g(1+{\rm e}^{-\beta E_\infty}){\rm e}^{-\beta E_\infty}}\, ,
\label{T_SP}
\end{eqnarray}
with $\ \beta\rightarrow 1/(k_B T_{SP})$.
This transcendental equation takes a simple form in some
limiting cases:

\begin{eqnarray}
  E_\infty=0:&\qquad& T_{SP}=\Delta_g^2/(\Delta_g+E_0)
\nonumber\\
  E_\infty=E_0:&\qquad& T_{SP}=\Delta_g e^{E_0/(k_BT_{SP})}
\label{cases}\\
  \Delta_g\rightarrow\infty:&\qquad& T_{SP}\rightarrow
                     \Delta_g-E_0+2E_\infty
\nonumber
\end{eqnarray}
For illustration we present in Fig.\ \ref{phases} (a)
$\ T_{SP}\ $ as a function of
the inter-chain coupling constant $\ \Delta_g\ $ for
$\ E_0=0.2\mbox{meV}$.
For $\ b(T_{SP})<0$ we obtain a first-order phase
transition. This is the case for values of $\ E_0\ $ 
larger than a certain critical value of the soliton energy
$\ E_c$, which is determined from $\ b(T_{SP})=0$ as

\begin{eqnarray}
E_c&=&E_\infty+\frac{\Delta_g^2}{T_{SP}}
\left(1+e^{E_\infty/T_{SP}}\right)^2
\nonumber\\&&\hspace{4ex}\cdot\,
\left(\sqrt{\frac{1+3e^{E_\infty/T_{SP}}+3e^{2E_\infty/T_{SP}}}
            {6\left(1+e^{E_\infty/T_{SP}}\right)^2}}
-\frac{1}{2}\right).
\label{secondorder}
\end{eqnarray}
For a fixed transition temperature $\ T_{SP}\ =\ 14.15\mbox{K}\ $ and
using Eq.~(\ref{T_SP}), $\ E_c\ $ and the corresponding inter-chain
coupling energy $\ \Delta_g\ $ can be calculated as a function of $\
E_\infty$. The resulting phase diagram is given in Fig.\
\ref{phases}(b). The numerical results presented throughout this paper
are obtained within the second-order regime, indicated by the crosses
in Fig.\ \ref{phases}(b).

\section{Self-consistency equation}

The effective Hamiltonian Eq.\ (\ref{Ising}) contains three
free parameters, namely $\ E_\infty$, $E_0$ and $\Delta_g$.
We examine two scenarios. The first is the case of
$\ \alpha_c>\alpha\ $ with $\ E_\infty=0$. The second
is the case of $\ \alpha_c<\alpha\approx 0.35\ $ with
$\ E_\infty=0.15\mbox{meV}$ corresponding to a gap
in the disordered phase of $\ 2E_\infty=0.3\mbox{meV}\ $ \cite{Chitra}. 
For each case we consider a range of $\ E_0\ $ (see Fig.\
\ref{phases}(b)) within the second-order regime, 
$E_\infty\le E_0\le E_c$,

\begin{equation}\label{values1}
E_\infty=0\,\mbox{meV};\quad
E_0=0,\,0.1,\,0.2,\,0.277\,\mbox{meV}
\end{equation}
\begin{equation}
2E_\infty=0.3\,\mbox{meV};\quad
E_0=0.15,\,0.3,\,0.4,\,0.474\,\mbox{meV} 
\label{values2}
\end{equation}
The largest value of $\ E_0\ $ for each $\ E_\infty\ $
corresponds to $\ E_c(E_\infty)$, compare Fig.\ \ref{phases}(b).
The experimental transition temperature
$\ T_{SP}=14.15\mbox{K}\ $ of CuGeO$_3$ then
determines the coupling constant $\ \Delta_g$ \cite{note1}.

\begin{figure}[htb]
\setlength{\unitlength}{0.96\textwidth}
\begin{picture}(0.49,0.84)
\epsfxsize=0.48\textwidth
\put(0,0.42){\epsffile{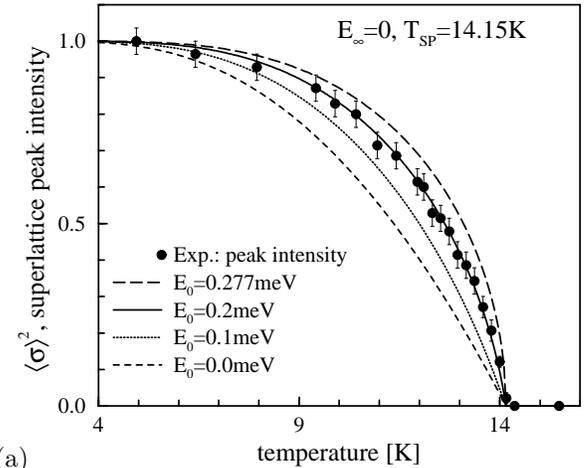}}
\epsfxsize=0.48\textwidth
\put(0,0){\epsffile{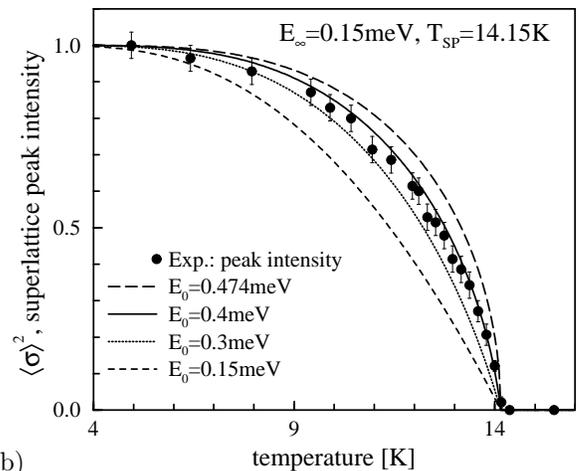}}
\put(0,0.44){(a)}
\put(0,0.02){(b)}
\end{picture}
\parbox{0.5\textwidth}{\caption{\label{distortion} \sl
The square of the spin-singlet order parameter 
($\langle\sigma\rangle^2$, lines) as a function of temperature,
for two different values of $E_\infty$ 
(soliton gap in the disordered phase)
and several values of $E_0$
($T=0$ soliton gap).
(a): $E_\infty=0$, i.e.\ $E_s=E_0\langle\sigma\rangle^2$. 
(b): $2E_\infty=0.3meV$, 
i.e.\ $E_s=E_\infty+(E_0-E_\infty)\langle\sigma\rangle^2$.
For comparison we plot the measured \protect\cite{Martin}
intensity of an additional super-lattice peak (filled circles).}}
\end{figure}

The order parameter $\ \langle\sigma\rangle\ $ is 
determined self-con\-sis\-tently as a function of temperature
by setting the derivative of the free energy with respect
to $\ \langle\sigma\rangle\ $ to zero,

\begin{equation}
\langle\sigma\rangle\ =\
\frac{\sinh(\beta\Delta_g\langle\sigma\rangle)}{
     2\frac{E_0-E_\infty}{\Delta_g\lambda_0}{\rm e}^{-2\beta E_s}+
\sqrt{\sinh^2(\beta\Delta_g\langle\sigma\rangle)
  +{\rm e}^{-2\beta E_s}}
                                                                 }
\mbox{.}\label{self}
\end{equation}
In Fig.\ \ref{distortion} we show the results for
$\ \sim\langle\sigma\rangle^2$ as a function of temperature
for the parameters given in (\ref{values1}) and (\ref{values2}).
We have also plotted the measured intensity of an
additional super-lattice peak \cite{Martin}, which is
proportional to the square of the lattice dimerization.
We have normalized  the experimental data such that agreement is
obtained in the low-temperature regime. 

The comparison between  
theory and the data for CuGeO$_3$ does not lead to a
determination of the frustration $\ \alpha\ $ but indicates
closeness to a first-order phase transition, as can be deducted from
the closeness of the experimental points in 
Fig.~\ref{distortion} to the critical curves where $\ E_0\ \approx\
E_c(E_\infty)$, compare Fig.~\ref{phases}(b). 
%
%

\section{Thermodynamics}

\begin{figure}[htb]
\setlength{\unitlength}{0.96\textwidth}
\begin{picture}(0.49,0.84)
\epsfxsize=0.48\textwidth
\put(0,0.42){\epsffile{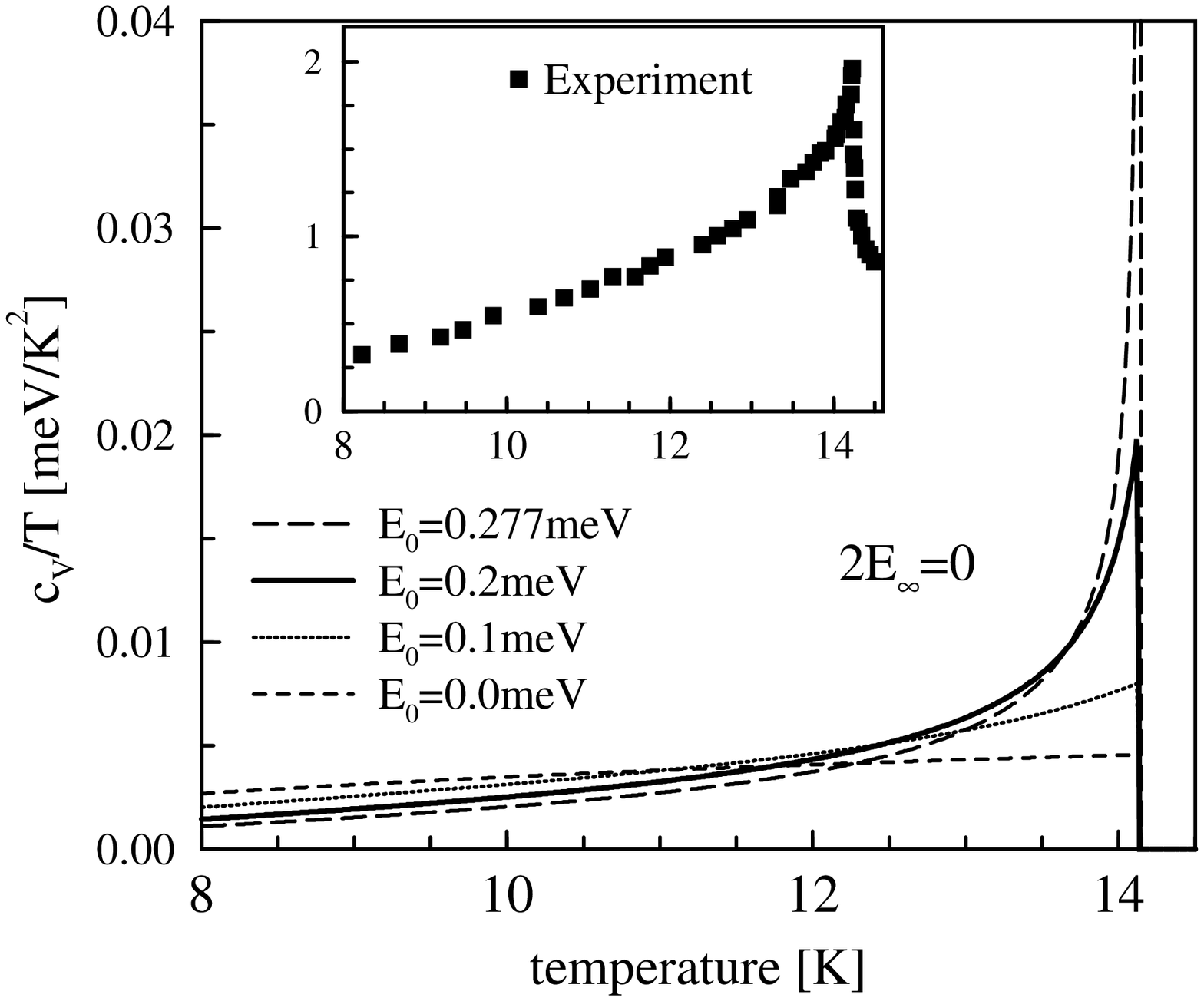}}
\epsfxsize=0.48\textwidth
\put(0,0){\epsffile{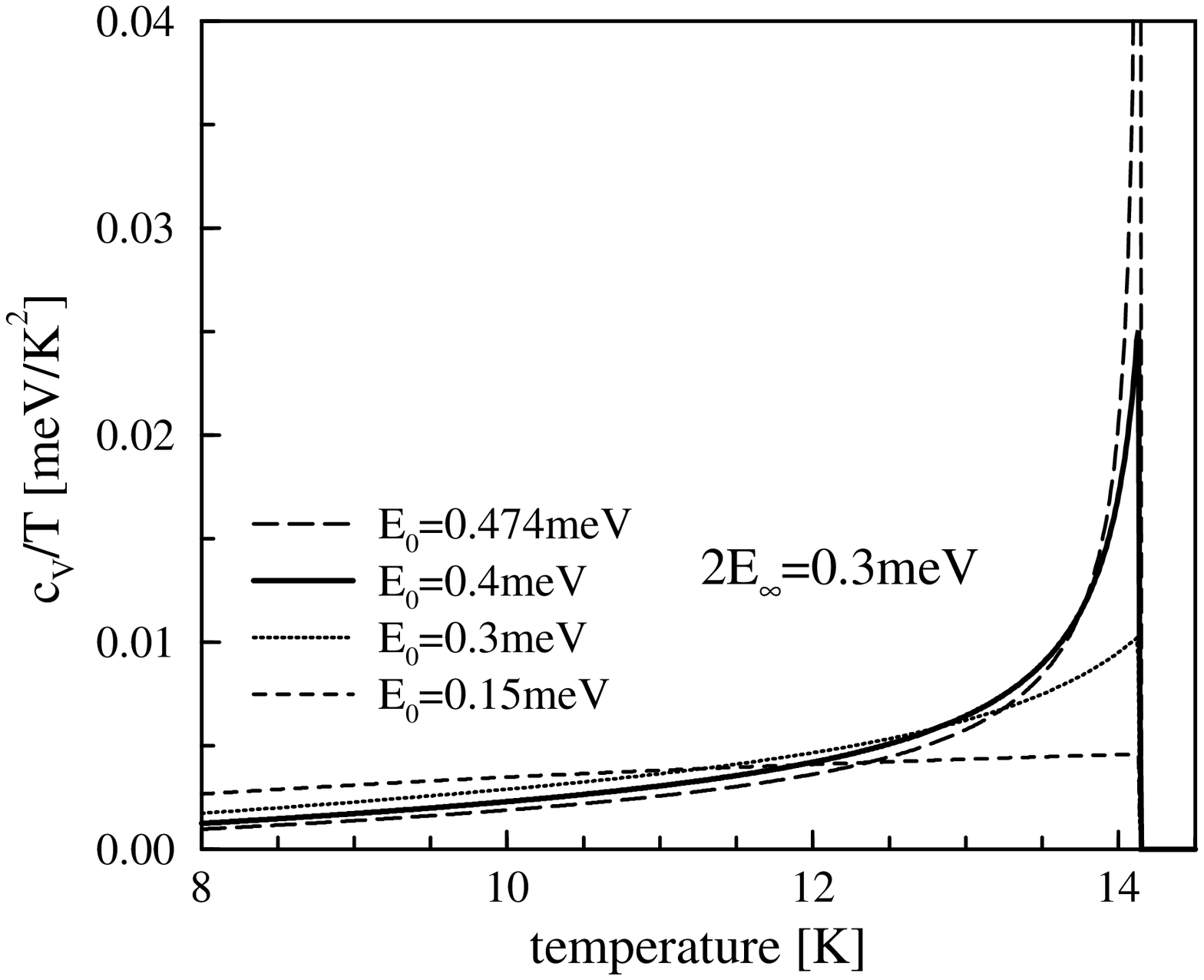}}
\put(0,0.44){(a)}
\put(0,0.02){(b)}
\end{picture}
\parbox{0.5\textwidth}{\caption{\label{cv} \sl
Specific heat for different parameters as a function of
temperature. (a): $E_\infty=0$. (b): $2E_\infty=0.3meV$. The inset in
graph (a) shows the experimental 
data \protect\cite{specific-heat} of $c_V/T$ in units of
$[\mbox{mJ}/\mbox{gK}^2]$ versus $T$
($10^{-2}\mbox{meV}/\mbox{K}^2$ corresponds to
$5.24\mbox{mJ}/\mbox{gK}^2$).}} 
\end{figure}

The entropy $\ S\ $ and the specific heat $\ c_V\ $
are obtained from the free energy (\ref{free}) via

\begin{equation}
S\ =\ -\frac{\partial F}{\partial T},\qquad
c_V\ =\ T\frac{\partial S}{\partial T}.
\label{c_V}
\end{equation}
In the disordered phase the entropy is  (for $E_\infty=0$)
temperature independent with
a value of $\ k_B\ln(2)/2\ $ per site, 
which is only half of the expected value for a spin $1/2$ chain.
This is due to the fact that we did neglect up to now 
the spin-degrees of freedom of the domain-walls. 
In Fig.\ \ref{cv} we present results for $\ c_V(T)\ $ for the
parameters given by (\ref{values1}) and (\ref{values2}). 
For small soliton excitation energies the results
are typically mean-field like. For values of $\ E_0\ $ approaching the
limit of the second-order phase regime the specific heat is strongly
enhanced near $T_{SP}$. It will diverge as the transition becomes
first-order. Note that the specific heat is linear in temperature for 
$\ E_\infty\le E_0 < E_c$, in the limit $\ T\rightarrow T_{SP}$ and
that the jump in the specific heat diverges as $\ E_0 \to E_c\ $ like
$\ (E_c - E_0)^{-1}$. Right at $\ E_0=E_c\ $ the specific heat
diverges like $\ (T_{SP}-T)^{-1/2}$. Note
that a similar divergence $\ \sim(T_{SP}-T)^{-0.4}\ $ has been
reported in an early measurement for CuGeO$_3$ \cite{Sahling} though
the exact value of the specific heat critical exponent for
CuGeO$_3$ is still controversial \cite{Liu,specific-heat}.

In the inset of Fig.\ \ref{cv} we present the
measured magnetic contribution to the specific heat of
CuGeO$_3$ \cite{specific-heat}. A straightforward
comparison with the results of the mean-field theory
is not possible since all the entropy of the effective Ising chain is
released in a mean-field approach. This corresponds to half of the
entropy of the spin chain. Experimentally only $\ \sim 10\%\ $ of the
magnetic entropy is released at $T_{SP}$ \cite{Weiden}, since
the exchange constant $\ J\approx 160\mbox{K}\gg T_{SP}$ and
the measured specific heat is consequently smaller
in magnitude than our mean-field result. 
The neglect of the soliton dispersion relation
(\ref{dispersion}) is, on a microscopic level, the reason for 
this discrepancy between theory and experiment. A qualitative
comparison is nevertheless possible and favors an $\ E_0\ $ close 
to the first-order phase transition.

\begin{figure}[htb]
\setlength{\unitlength}{0.96\textwidth}
\begin{picture}(0.49,0.84)
\epsfxsize=0.48\textwidth
\put(0,0.42){\epsffile{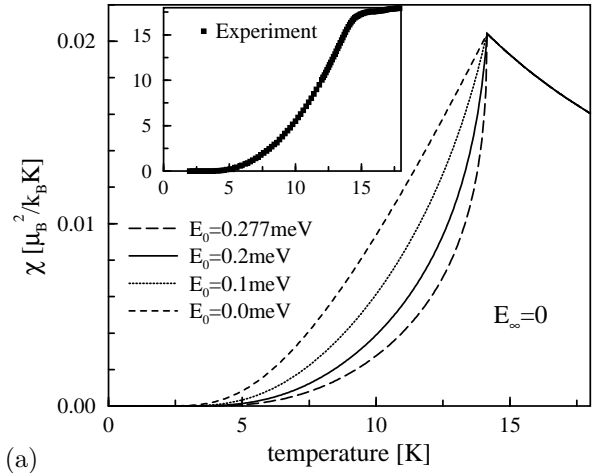}}
\epsfxsize=0.48\textwidth
\put(0,0){\epsffile{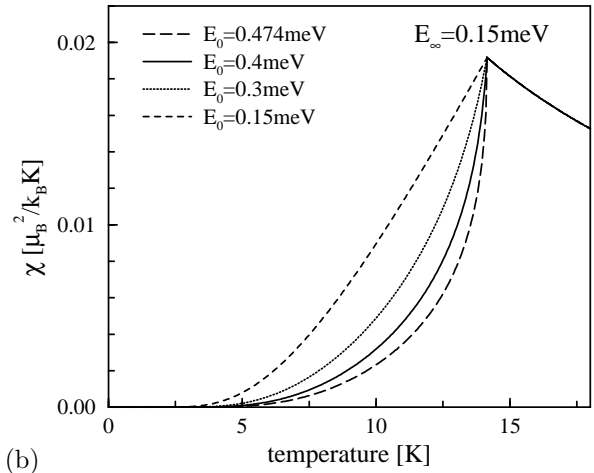}}
\put(0,0.44){(a)}
\put(0,0.02){(b)}
\end{picture}
\parbox{0.5\textwidth}{\caption{\label{chi} \sl
Magnetic susceptibility for different parameters as a function of
temperature. (a): $E_\infty=0$. (b): $2E_\infty=0.3meV$. The inset
in graph (a) shows the experimental data \protect\cite{chi} of
$\chi$ in units of $[10^{-9}\mbox{m}^3/\mbox{mole}]$ versus $T$ 
($0.01\mu_B^2/k_B\mbox{K}$ correspond to $47.12\cdot
10^{-9}\mbox{m}^3/\mbox{mole}$).}} 
\end{figure}

Up to know we did not take the spin-degree of freedom of the
solitons into account, as they just contribute a constant
factor to the partition function in the paramagnetic case.
As we have no magnetic interactions between the spins of
different solitons in our model we can evaluate the
magnetic susceptibility simply by Curie's law

\begin{equation}
\chi(T)=\frac{g^2\mu_B^2S(S+1)}{3k_BT}\;n(T)\approx
1.16\frac{\mu_B^2}{k_BT}\;n(T),
\label{Curie}
\end{equation}
where $\ \mu_B=e\hbar/(2m_ec)\ $ is the Bohr magneton, $\ g=2.15\ $ the
measured \cite{chi}
g-factor of the $Cu^{2+}-$ion, $\ S=1/2\ $ and
$\ n(T)\ $ the density of thermally activated solitons per site.
$\ n(T)\ $ is obtained differentiating the free energy
with respect to $E_s$:

\begin{equation}
n(T)\ =\ \frac{\partial F/L}{\partial E_s}=\frac{1}{2\lambda_0}
     \frac{e^{-2\beta E_s}}
     {\sqrt{\sinh^2(\beta\Delta_g\langle\sigma\rangle)+e^{-2\beta E_s}}}
       \mbox{.}
\label{n(T)}
\end{equation}
Above $T_{SP}$ this expression reduces to

\begin{equation}
n(T>T_{SP})=\frac{1}{2}\frac{1}{e^{\beta E_\infty}+1}\mbox{.}
\label{n_T}
\end{equation}
The mean number of solitons attains $1/4$ per site,
as the temperature goes to infinity, corresponding to half a
soliton per dimer. The results for the magnetic susceptibility
are shown in Fig.\ \ref{chi}. The fast drop of
$\ \chi(T)\ $ below $T_{SP}$ for larger values of the 
soliton excitations energies is again reminiscent of the experimental
data for CuGeO$_3$ \cite{chi}. Our susceptibility rises though much
higher at $\ T_{SP}\ $ than the experimental data which is a direct
consequence of the neglected soliton dispersion (\ref{dispersion}).
In the limit of large temperatures $\ T\ \gg J\
$ the theoretical curve drops to $1/4$ of the experimental
points as a consequence of the aforementioned soliton 
density (\ref{n_T}) (at $\ 300\mbox{K}\approx\ 2J$ 
it has dropped to about $1/2$ of the magnitude
of the experimental data). 

Note that for the curves plotted as solid lines in
Fig.~\ref{distortion}, Fig.~\ref{cv}, and Fig.~\ref{chi}, which are
closest to the experimental data within the parameters chosen by us,
entropy, specific heat, and susceptibilty are overestimated by about
the same factor of five.   
%
%

\section{Singlet--Triplet Gap}

\begin{figure}[htb]
\setlength{\unitlength}{0.96\textwidth}
\begin{picture}(0.49,0.84)
\epsfxsize=0.48\textwidth
\put(0,0.42){\epsffile{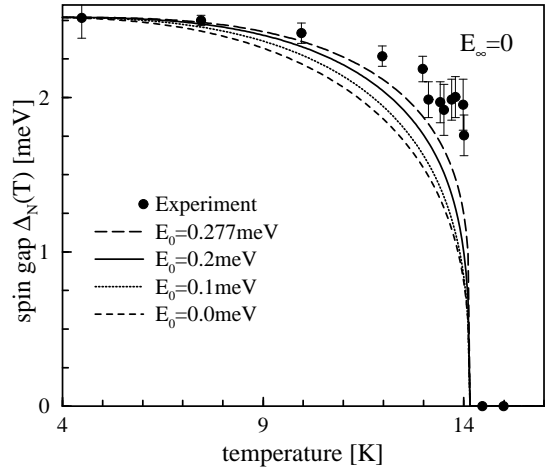}}
\epsfxsize=0.48\textwidth
\put(0,0){\epsffile{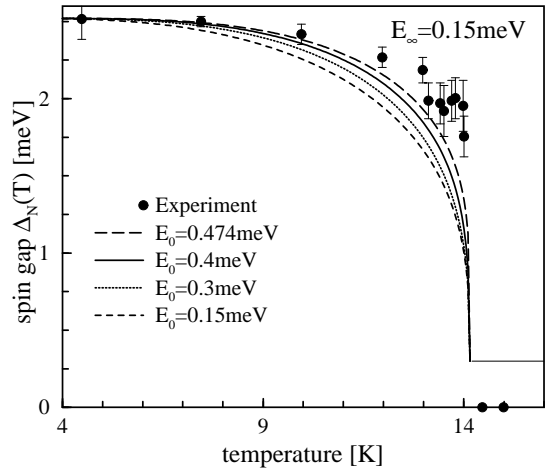}}
\put(0,0.44){(a)}
\put(0,0.02){(b)}
\end{picture}
\parbox{0.5\textwidth}{\caption{\label{gap} \sl
The gap to triplet excitations for several parameters
as a function of temperature. 
(a): $E_\infty=0$. (b): $2E_\infty=0.3meV$.}}
\end{figure}

The gap to triplet excitations is given by Eq.\ (\ref{Delta_N}),

\begin{equation}
\Delta_N\ =\ 2\left(E_\infty +(E_0-E_\infty)\langle\sigma\rangle^2\right)
  + c^\prime\,J\,\left({\Delta_g\langle\sigma\rangle\over J}
                \right)^{2/3},
\label{triplet_gap}
\end{equation}
with $\ c^\prime=2.33$. A straightforward application of
(\ref{triplet_gap}) would yield, compared with experiment,
a much too large zero-temperature gap
$\ 2E_0+c^\prime J(\Delta_g/J)^{2/3}$. This is so since the
order parameter $\ \langle\sigma\rangle(T=0)\ $ takes the
value one in the molecular-field approximation, while it is
much smaller than unity for CuGeO$_3$. We have therefore decided
to use the parameter $\ c^\prime\ $ in (\ref{triplet_gap}) to
fix $\ \Delta_N(T=0)\ $ to the experimentally observed value
$\ 2.5\mbox{meV}$. The results are given in
Fig.\ \ref{gap}, together with the measured gap for
CuGeO$_3$ \cite{Martin}.
%
%

\section{Conclusions}

We have discussed a simple mean-field theory for 
spin-Peierls transitions applicable both for
phonon-driven ($\alpha<\alpha_c$) and for
magnetically driven ($\alpha>\alpha_c$) spin-Peierls
transitions. We have applied the approach to
CuGeO$_3$ and found that it is not possible to
determine uniquely from the experimentally measured
temperature dependence of the order-parameter
the magnitude of the frustration parameter $\alpha$.

The theory allows both for a first-order and a second-order
spin-Peierls transition, depending on the parameters of
the model. We find that the parameters which fit
experiments best indicate that CuGeO$_3$ is
close to a first-order phase transition.

We would like to thank P. Lemmens and B. B\"uchner 
for discussions and J.~Stolze for pointing out a straightforward
derivation of the $\ \delta^{2/3}\ $ scaling law. The support of the
DFG is gratefully acknowledged.


\begin{thebibliography}{99}
\bibitem{note1}  The complete sets of parameter values
                  $(E_0,E_\infty,\Delta_g)$ are in [meV]:
                  (0,0,1.22),
                  (0.1,0,1.31), (0.2,0,1.39), (0.277,0,1.45)
                  for $\alpha<\alpha_c$ and
                  (0.15,0.15,1.05), (0.3,0.15,1.2),
                  (0.4,0.15,1.28), (0.474,0.15,1.33) for
                  $\alpha>\alpha_c$.
\end{thebibliography}
\end{document}